\documentclass[twocolumn,superscriptaddress,amsfont,amssymb,amsmath, showpacs,balancelastpage, nofootinbib]{revtex4-1}
\usepackage{graphicx,longtable,color,array}
\usepackage[hidelinks]{hyperref}
\usepackage[usenames,dvipsnames]{xcolor}
\usepackage{amssymb,mathrsfs}
\usepackage{mathtools}
\usepackage{bm}
\usepackage{amsmath}
\usepackage[caption=false,justification=justified]{subfig}
\usepackage[compat=1.1.0]{tikz-feynman}
\usepackage{booktabs,longtable,ctable,multirow}
\usepackage{exscale,relsize}
\usepackage[normalem]{ulem}
\usepackage{enumerate}
\usepackage{colortbl} 
\usepackage[dvipsnames]{xcolor} 
\usepackage{pifont} 
\usepackage{adjustbox}

\allowdisplaybreaks

\usepackage{tikz}
\usepackage[compat=1.1.0]{tikz-feynman}

\newcommand{\Gprop}{
\begin{tikzpicture}[baseline]
\begin{feynman}
\vertex  [label=90:$\scriptstyle{\mu\nu}$](a);
\vertex [right=1.5cm of a, label=90:$\scriptstyle{\rho\sigma}$] (b);
\diagram* {
(a) -- [gluon, edge label=$k$] (b)
}; 
\end{feynman}
\end{tikzpicture}
}

\newcommand{\FRulezero}{
\begin{tikzpicture}[baseline]
\begin{feynman}
\vertex [dot, label=90:$\tau_a$] (a) {};
\vertex [right=0.8cm of a] (b);
\diagram* {
(a) -- [gluon, momentum={[arrow shorten=0.3pt]$k$}, insertion=1] (b),
}; 
\end{feynman}
\end{tikzpicture}}

\newcommand{\FRuleone}{
\begin{tikzpicture}[baseline]
\begin{feynman}
\vertex [ dot, label=90:$\tau_a$] (a) {};
\vertex [below=1.2cm of a, dot] (c) {};
\vertex [below right=0.8cm of a] (b);
\diagram* {
(c) -- [gluon] (a),
(a) -- [gluon, momentum={[arrow shorten=0.3pt]$k$}, insertion=1] (b)
}; 
\end{feynman}
\end{tikzpicture}
}

\newcommand{\VTidal}{
\begin{tikzpicture}[baseline]
\begin{feynman}
\vertex [square dot, label=180:$\tau_a $](z){};
\vertex [right=0.6cm of z](c);
\vertex [above=0.5cm of c, label=45:$\scriptstyle{\kappa\lambda}$] (b);
\vertex [below=0.5cm of c, label=315:$\scriptstyle{\mu\nu}$] (a);
\diagram* {
(z) -- [gluon, momentum'={[arrow shorten=0.2pt]$\ell_1$}, insertion=1] (a),
(z) -- [gluon, momentum={[arrow shorten=0.2pt]$\ell_2$}, insertion=1] (b),
}; 
\end{feynman} 
\end{tikzpicture}}

\newcommand{\MdiagT}{
\begin{tikzpicture}[baseline]
\begin{feynman}
\vertex [ dot] (a) {};
\vertex [below=2cm of a, dot] (b) {};
\vertex [right=1cm of a] (w);
\vertex [below=2cm of w] (u);
\vertex [right=2cm of a,  square dot,  label=90:$c_{X_1^2}$] (a') {};
\vertex [below=2cm of a', dot] (b') {};
\diagram* {
(b) -- [gluon] (a),
(a) -- [gluon, out=315, in=225] (a'),
(b') -- [gluon] (a'),
(w) -- [scalar] (u)
}; 
\end{feynman}
\end{tikzpicture}
}

\newcommand{\NdiagT}{
\begin{tikzpicture}[baseline]
\begin{feynman}
\vertex [ dot] (a) {};
\vertex [below=2cm of a, dot] (b) {};
\vertex [right=2cm of a, square dot,  label=90:$ c_{X_1^2}$] (a') {};
\vertex [below=2cm of a', dot] (b') {};
\vertex [right=1cm of b] (w);
\vertex [right=1cm of a] (u);
\diagram* {
(b) -- [gluon ] (a),
(a') -- [gluon] (b),
(b') -- [gluon] (a'),
(u) -- [scalar] (w) 
}; 
\end{feynman}
\end{tikzpicture}
}

\newcommand{\IYdiagT}{
\begin{tikzpicture}[baseline]
\begin{feynman}
\vertex [square dot, label=90:$c_{X_1^2}$] (a) {};
\vertex [below=2cm of a, dot] (b) {};
\vertex [left=2cm of b, dot] (b') {};
\vertex [above=1cm of b'] (c');
\vertex [above=2cm of b', dot] (a') {};
\vertex [left=1cm of b] (w);
\vertex [left=1cm of a] (u);
\diagram* {
(b) -- [gluon] (a),
(a) -- [gluon] (c'),
(b') -- [gluon] (c') -- [gluon] (a'),
(u) -- [scalar] (w) 
}; 
\end{feynman}
\end{tikzpicture}}

\newcommand{\be}{\begin{equation}}      
\newcommand{\ee}{\end{equation}}

\newcommand{\mpl}{m_{\text{Pl}}}

\newcommand{\vv}{{\bf v}}
\newcommand{\vb}{{\bf b}}

\newcommand{\vn}{{\bf n}}
\newcommand{\ve}{{\bf e}}

\def\dd{\delta\!\!\!{}^-\!}
\newcommand{\uu}{{\cal U}}

\newcommand{\firstc}{\tikz{\draw[line width=0.5pt] (0,-5pt) -- (0,0pt) -- (3pt,-2pt) ;}}
\newcommand{\secondc}{\tikz{\draw[line width=0.5pt] (0,0pt) -- (0,-5pt) -- (3pt,-2pt) ;}}

\newcommand{\tabeq}[2]{ \parbox{#1}{  \be\begin{aligned}#2 \end{aligned} \nonumber \ee }}

\begin{document}

\title{Gravitational Bremsstrahlung with tidal effects\\ in the post-Minkowskian expansion}
\author{Stavros Mougiakakos}  
\affiliation{Institut de physique th\' eorique, Universit\'e  Paris Saclay CEA, CNRS, 91191 Gif-sur-Yvette, France}
\affiliation{Laboratoire Univers et Th\' eories, Observatoire de Paris, Universit\'e PSL, Universit\'e Paris Cité, CNRS, F-92190 Meudon, France}
\author{Massimiliano Maria Riva} 
\author{Filippo Vernizzi}
\affiliation{Institut de physique th\' eorique, Universit\'e  Paris Saclay CEA, CNRS, 91191 Gif-sur-Yvette, France}

\begin{abstract}

We compute
the mass and current quadrupole tidal corrections to the four-momentum and energy flux radiated during the scattering of two spinless bodies, at leading order in $G$ and at all orders in the velocities, using the effective field theory worldline approach. 
In particular, we derive  the conserved stress-energy tensor linearly coupled to gravity generated by the two bodies, including tidal fields, and the waveform in direct space.
The integral is solved using scattering amplitude techniques. We show that our expressions are consistent with existing results up to the next-to-next-to-leading order   in the post-Newtonian expansion.

\end{abstract}

\maketitle


{\em Introduction --} The direct detection of gravitational waves from binary black holes \cite{Abbott:2016blz} and neutron stars \cite{TheLIGOScientific:2017qsa} has opened  an new way  to test gravity in the strong-field regime \cite{Berti:2015itd} and explore fundamental physics \cite{Barausse:2020rsu}. An important target of current and future observations is the measurement of tidal deformations during the coalescence of compact objects \cite{Damour:1992qi,Damour:1993zn,Goldberger:2005cd,Hinderer:2007mb,Flanagan:2007ix,Damour:2009vw,Binnington:2009bb,Hinderer:2009ca,Kol:2011vg,Damour:2012yf,Favata:2013rwa},  which may shed light on  the internal structure of neutron stars \cite{Baiotti:2016qnr}, the nature of black holes \cite{Barack:2018yly} or the existence of more exotic astrophysical objects \cite{Buonanno:2014aza,Cardoso:2019rvt,Baumann:2019ztm}.

Tidal deformations affect 
the conservative two-body dynamics as well as the emitted energy in gravitational waves. They
have been studied utilising different analytical 
techniques,
most notably the post-Newtonian (PN) 
expansion
\cite{Mora:2003wt,Vines:2010ca,Vines:2011ud,Henry:2019xhg,Henry:2020pzq,Henry:2020ski},  the effective-one-body approach \cite{Damour:2009wj,Bini:2012gu,Steinhoff:2016rfi}, Non-Relativistic-General-Relativity (NRGR) \cite{Goldberger:2004jt,Goldberger:2007hy,Foffa:2013qca,Rothstein:2014sra,Porto:2016pyg,Levi:2018nxp} and the self-force formalism \cite{Bini:2014zxa, Bini:2018svh, Bini:2018kov,Bini:2018dki} (see \cite{Dietrich:2020eud} for a review).

Another 
technique that has been employed to 
study the gravitational two-body 
problem is the post-Minkowskian (PM) method
\cite{Bertotti:1956pxu,Bertotti:1960wuq,Havas:1962zz,Westpfahl:1979gu,Portilla:1980uz,Bel:1981be,Westpfahl:1985tsl,Damour:2016gwp,Damour:2017zjx}, consisting in expanding the gravitational dynamics for small  interactions, while keeping the velocities fully relativistic. 
It  has been recently  subject of great  interest and activity, in particular in association with  the effective-one-body approach \cite{Damour:2016gwp,Damour:2017zjx,Bini:2017xzy,Bini:2018ywr,Vines:2018gqi,Damour:2019lcq,Antonelli:2019ytb},  scattering amplitude techniques \cite{Neill:2013wsa,Bjerrum-Bohr:2013bxa,Luna:2017dtq,Bjerrum-Bohr:2018xdl,Kosower:2018adc,Cheung:2018wkq,Bern:2019nnu,Bern:2019crd,Cristofoli:2019neg,Cristofoli:2020uzm,Cheung:2020gyp,Bern:2020buy,Bern:2021dqo,Bern:2021yeh,DiVecchia:2021bdo,Bjerrum-Bohr:2021vuf,Bjerrum-Bohr:2021din, Damgaard:2021ipf}, and worldline approaches \cite{Foffa:2013gja,Goldberger:2016iau,Goldberger:2017vcg,Kalin:2020mvi,Kalin:2020fhe,Loebbert:2020aos,Mogull:2020sak,Liu:2021zxr,Cho:2021mqw,Dlapa:2021npj,Dlapa:2021vgp,Jakobsen:2021lvp,Jakobsen:2021zvh,Cho:2022syn,Jakobsen:2022fcj}. Tidal effects have been studied with the PM expansion in \cite{Bern:2020uwk,Cheung:2020sdj,AccettulliHuber:2020oou,Haddad:2020que,Aoude:2020onz,Bini:2020flp,Kalin:2020lmz,Cheung:2020gbf,Huber:2020xny,Goldberger:2019xef,Goldberger:2020wbx,Goldberger:2020fot}. These developments concern the scattering of  
two bodies moving on unbounded orbits but computed observables can be extended to the case of bound orbits 
by applying the so-called ``boundary-to-bound'' (B2B) dictionary, consisting in an  analytic continuation between hyperbolic and elliptic motion \cite{Kalin:2019rwq,Kalin:2019inp,Bini:2020hmy,Cho:2021arx}.

A long-standing and, until recently, unsolved problem was the calculation of the    four-momentum radiated in gravitational waves---the so-called gravitational Bremsstrahlung---during the scattering of two spinless bodies, at leading PM order, i.e.~at ${\cal O}(G^3)$. 
This was finally obtained very recently in \cite{Herrmann:2021lqe,Herrmann:2021tct} via the amplitude-based method of \cite{Kosower:2018adc}, in \cite{DiVecchia:2021bdo} using the eikonal approach 
and in \cite{Riva:2021vnj} by a classical effective field theory (EFT) worldline  approach. (See also \cite{Amati:1990xe, DiVecchia:2019myk, DiVecchia:2019kta, Bern:2020gjj, DiVecchia:2020ymx, Huber:2020xny,Damour:2020tta,DiVecchia:2021ndb,Bini:2021gat,Jakobsen:2021smu,Mougiakakos:2021ckm} for previous work on radiation effects. Earlier pioneering studies include \cite{Peters:1970mx,Thorne:1975aa,Crowley:1977us,Kovacs:1977uw,Kovacs:1978eu,Turner:1978zz,Westpfahl:1985tsl}. 
Moreover, see \cite{Saketh:2021sri,Bern:2021xze} for conservative and radiative effects  in QED.) 
Crucially, these calculations strongly benefited from several computational tools developed in the high-energy community \cite{Parra-Martinez:2020dzs}, such as reduction to master integrals   by Integration-by-Parts (IBP) identities \cite{Tkachov:1981wb,Chetyrkin:1981qh, Smirnov:2012gma} and   differential equations \cite{Kotikov:1990kg,Bern:1992em,Gehrmann:1999as,Henn:2013pwa} to solve the latter using the near-static regime as initial conditions.

In particular, in \cite{Riva:2021vnj} two of us showed that it is possible to use these tools to directly compute radiated observables in the PM expansion without going through the classical limit of scattering amplitudes.
Indeed, the  emitted four-momentum was obtained by phase-space integration of the graviton momentum weighted by the modulo squared of the classical radiation amplitude \cite{Jakobsen:2021smu,Mougiakakos:2021ckm}, the latter being derived by matching to the conserved stress-energy tensor linearly coupled to gravity, generated by localized sources. The phase-space integral was then recasted as a 2-loop integral that we solved with the aforementioned techniques. 

In this letter we use the same approach but we go beyond 
the minimally coupled case, and we compute 
for the first time the effect of tidal deformations on the  four-momentum radiated into gravitational waves during the scattering of the two bodies.   From this, extending the technique recently developed in \cite{Cho:2021arx},   we also compute the tidal corrections to the emitted energy flux, which is valid for both open and closed orbits.
We focus on the leading tidal contributions to the orbital dynamics, i.e.~to quadrupolar deformations, but the extension to higher multipoles can be straightforwardly obtained using the same approach.

The article is organized as follows. We first  define the Feynman rules in the case of tidal couplings, which will allow us to derive the stress-energy tensor linearly coupled to gravity, and the waveform in direct space, at leading PM order. From the stress-energy tensor, we compute, using reverse unitarity, the  total four-momentum radiated into gravitational waves,  and from this  the emitted flux. We then use the B2B dictionary \cite{Kalin:2019rwq,Kalin:2019inp,Bini:2020hmy,Cho:2021arx} to check our results   
with PN derivations \cite{Henry:2020ski}.


{\em Leading PM tidal effects --} We consider the scattering of two gravitationally interacting  spinless bodies with mass $m_1$ and $m_2$, approaching each other from infinity.
Using the mostly minus metric signature, setting $\hbar=c=1$ and defining the Planck mass as $\mpl \equiv  1/\sqrt{32 \pi G}$, the total action describing the dynamics with tidal effects reads 
\be
\label{eq:SAction}
S =   - 2 \mpl^2 \int d^4 x \sqrt{-g} R  + S_{\rm pp} +S_{\text{tidal}}\; .
\ee
At leading order in their size, the bodies are described by point-particle  actions, 
\be
\label{Actionpp}
S_{\rm pp} =  - \sum_{a=1,2} \frac{m_a}{2}\int d\tau_a  g_{\mu\nu}(x_a) \uu_a^\mu(\tau_a) \uu_a^\nu(\tau_a) \;,
\ee 
where  $\tau_a$  and $\uu_a^\mu \equiv \nolinebreak {d x_a^\mu}/{d \tau_a}$ (with $ {\cal U}^a_\mu {\cal U}_a^\mu = 1$)
are, respectively, the proper time and the four-velocity of body $a$.
Note that we have used the Polyakov-like form of the action and fixed the {\em einbein} to unity, which  simplifies the gravitational coupling to the matter sources \cite{Galley:2013eba,Kuntz:2020gan,Kalin:2020mvi}.

Tidal effects are included  by augmenting the
point-particle action  with non-minimal worldline couplings involving higher-order derivatives of the gravitational field \cite{Goldberger:2004jt}. At leading PM order, 
only {\em linear} tidal deformations, i.e., those  whose response is linear in the external gravitational field, are relevant. These are described by 
couplings quadratic in the Weyl tensor $C_{\mu\alpha\nu\beta}$ evaluated at the particle position.
The Weyl tensor   can be decomposed in terms of the gravito-electric and gravito-magnetic fields, defined as
\begin{align}
\label{EB}
E_{\mu\nu} \equiv C_{\mu\alpha\nu\beta}\uu^{\alpha}\uu^{\beta}\, , \quad B_{\mu\nu} \equiv  \frac{1}{2}\epsilon_{\alpha\beta\gamma\mu}C^{\alpha\beta}_{\ \ \ \delta\nu}\uu^{\gamma}\uu^{\delta}\, ,
\end{align}
where $\epsilon_{\alpha\beta\gamma\mu}$ is the Levi-Civita tensor. At lowest-order in derivatives, and restricting to  parity-even operators for symmetry reasons, the action describing tidal deformations is given by 
\be
\label{linear}
S_{\text{tidal}}=\sum_{a=1,2}\int d\tau_a  \Big(c_{E_a^2} E^a_{\mu \nu}E_a^{\mu \nu} +c_{B_a^2} B^a_{\mu \nu} B_a^{\mu \nu}\Big) \, ,
\ee
where $c_{E_a^2}$ and $c_{B_a^2}$ are  Wilson coefficients related to the relativistic Love numbers $k^{(2)}_a$ and $j^{(2)}_a$ \cite{Bini:2012gu}, respectively as  $c_{E_a^2} = \frac{1}{6} k^{(2)}_a R^5_a /G$, $c_{B_a^2} = \frac{1}{32} j^{(2)}_a R^5_a /G$, with $R_a$ the radius of the object $a$.
Tidal operators can be equally defined by replacing the the Weyl tensor in eq.~\eqref{EB} with the Riemann tensor:  the difference can be   removed by field redefinitions, see e.g.~\cite{Dixon:1975si,Goldberger:2004jt,Henry:2019xhg}. Here we will use the  Riemann tensor because it leads to simpler calculations.
In full generality, one could also add to eq.~\eqref{linear} operators  including spatial derivatives, orthogonal to the worldline of the body, of the gravito-electric or gravito-magnetic field, as well as time derivatives along the worldline \cite{Bini:2012gu}.  Higher spatial derivatives describe higher-order multipolar deformations of the objects while  time derivatives account for the time dependence of the Wilson coefficients, see e.g.~\cite{Goldberger:2005cd, Porto:2016pyg}

Following \cite{Mougiakakos:2021ckm,Riva:2021vnj}, our first goal is to compute the stress-energy tensor $T^{\mu\nu}$ defined as the linear term sourcing the gravitational field  in the effective action \cite{DeWitt:1967ub,Abbott:1981ke,Goldberger:2004jt}, i.e.,
\be
\label{bfea}
\Gamma [x_a, h_{\mu \nu} ] = - \frac{1}{2 \mpl} \int d^4 x  \,T^{\mu \nu} (x) h_{\mu \nu} (x) \;,
\ee
with $h_{\mu\nu} \equiv \mpl (g_{\mu\nu} - \eta_{\mu\nu})$, which includes  contributions  from both the 
bodies 
and the gravitational self-interactions.
To do so, we use a matching procedure consisting in expanding  the action \eqref{eq:SAction} for small $h_{\mu\nu}$ and computing perturbatively all Feynman diagrams with one external graviton.  The stress-energy tensor is obtained by matching this result with the one computed using eq.~\eqref{bfea}. 
To proceed,
we need to introduce the Feynman rules. 

Adding the usual de Donder gauge-fixing term to eq. (\ref{eq:SAction}),
 from the quadratic part of the gravitational action one can extract the graviton propagator, 
\be 
\Gprop = \frac{i}{k^2} \mathbb{P}_{\mu \nu \rho\sigma}    \; ,
\ee
where 
$\mathbb{P}_{\mu \nu\rho\sigma} \equiv  \eta_{\mu(\rho}\eta_{\sigma)\nu}-\frac{\eta_{\mu\nu}\eta_{\rho\sigma}}{2} $.
(The boundary conditions that specify the contour of integration in the complex $k^0$-plane  are discussed in \cite{Mougiakakos:2021ckm}.)
Furthermore, expanding the Einstein-Hilbert action in \eqref{eq:SAction} at cubic order we can extract the cubic graviton vertex.

We also need to find the Feynman rules coming from the interaction of
gravity with the external sources, 
i.e.~the two bodies. 
These are of two types: minimal and tidal. For the former, from eq.~\eqref{Actionpp} one sees that there is only one linear interaction vertex.
As discussed in \cite{Mougiakakos:2021ckm} (see also \cite{Kalin:2020mvi, Kalin:2020fhe}),
we isolate the powers of $G$ by expanding the position and velocity of the bodies around straight trajectories, i.e., 
\begin{align}
x_a^\mu(\tau_a) & = b^\mu_a + u^\mu_a \tau_a + \delta^{(1)} x_a^\mu(\tau_a) +\dots \; ,\\
\uu_a^\mu (\tau_a) & = u_a^{\mu} + \delta^{(1)} u_a^\mu (\tau_a) +\dots \; ,
\end{align}
where $ u_a$ is the (constant) asymptotic incoming velocity and $b_a$ is the body displacement orthogonal to  it, $b_a \cdot u_a =\nolinebreak 0$, while  $\delta^{(1)} x_a^\mu$ and $ \delta^{(1)} u_a^\mu$ are respectively the deviation from the straight trajectory and constant velocity of body $a$ at order $G$, induced by the gravitational interaction. 
With this expansion we obtain the usual Feynman rules for the leading and next-to-leading PM-order graviton coupling in the point-particle  case \cite{Mougiakakos:2021ckm}, respectively represented by the diagrams  
\be 
\FRulezero  \;, \qquad \raisebox{15pt}{\FRuleone} \;,
\ee
where a filled dot denotes a minimally-coupled particle evaluated using the straight worldline and the cross attached to the wiggly line is there to remind us that there is no propagator attached to the straight worldline.  Their explicit expressions can be found in \cite{Mougiakakos:2021ckm,Riva:2021vnj}.

Moreover, we need to provide the Feynman rules from tidal contributions. In this case, from eq.~\eqref{linear} there is no tidal coupling linear in the graviton.
Tidal couplings of two gravitons to the 
body 
can be 
directly computed from the action using that
\be
\begin{split}
\mathcal{M}^{E_a}_{\mu\nu \alpha\beta}(\ell) \equiv & \ \frac{2 \delta E^a_{\mu \nu} }{\delta h^{\alpha \beta}(\ell)} =   \eta_{\mu \sigma} \eta_{\nu \rho} u_{a}^{\sigma}u_{a}^{\rho} \ell_{\alpha}\ell_{\beta}\\
& +(\ell\cdot u_a)^2\eta_{\alpha(\mu}\eta_{\nu)\beta} -2 (\ell \cdot u_a)u_a^\rho \eta_{\rho (\mu}\eta_{\nu)(\alpha} \ell_{\beta)} \;, \\
\mathcal{M}^{B_a}_{\mu\nu\alpha\beta}(\ell) \equiv & \ \frac{2 \delta B^a_{\mu \nu} }{\delta h^{\alpha \beta}(\ell)} =   \frac{1}{2} l^\rho u_a^\sigma \epsilon_{\rho \sigma \alpha(\mu}\big[\eta_{\nu)\beta}(\ell \cdot u_a) \\
&  - \eta_{ \nu) \rho } u_a^\rho\ell_{\beta}\big] + (\alpha \leftrightarrow \beta) \;,
\end{split}
\ee
where we use the flat metric  $\eta_{\mu \nu}$ to raise and lower indices.
At leading PM order one obtains
\be 
\label{Tvertex}
\VTidal  \!\!\!\!\!  = -\frac{i \delta^2 S_{\rm tidal}}{\delta h^{\mu \nu} (\ell_1) \delta h^{\kappa \lambda} (\ell_2) } \equiv V_{\mu\nu,\kappa\lambda} (\ell_1,\ell_2) \;,
\ee
where
\be 
V_{\mu\nu,\kappa\lambda}  = i \!\!\!\! \sum_{X=E, B }  \sum_{a=1,2}  \frac{c_{X_a^2} }{4 \mpl^2}   \int d\tau_a e^{i (\ell_1+\ell_2) \cdot (b_a+u_a \tau_a)}   \Pi^{{X}_a}_{\mu\nu,\kappa\lambda}  \;,
\ee
with
\be 
\label{PiXa}
\Pi^{ X_a}_{\mu\nu,\kappa\lambda}(\ell_1,\ell_2) \equiv \mathcal{M}^{X_a}_{\mu\nu\alpha\beta}(\ell_1) {\mathcal{M}^{X_a}_{\kappa\lambda}}^{\alpha\beta}(\ell_2) \;.
\ee
On the left-hand side of eq.~\eqref{Tvertex}, 
the square denotes a tidally-coupled particle evaluated using the straight worldline.
We have verified that our expression agrees with that 
that 
can be read off  from  the 4-point amplitude  at leading PM order obtained in Ref.~\cite{Bern:2020uwk}.


\begin{figure}[t!!]
\includegraphics[width=0.48\textwidth]{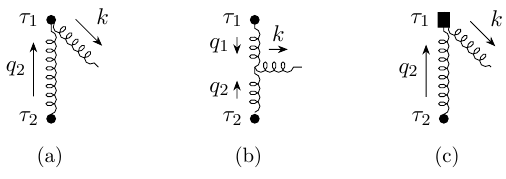}
\caption{The  Feynman diagrams needed for the computation of the stress-energy tensor: (a) and  (b) are the point-particle contributions, and (c) is the tidal one. The symmetric terms are obtained by exchange of $1 \leftrightarrow 2 $.}
\label{Figure}
\end{figure}
{\em Stress-energy tensor with tidal effects --} The  stress-energy tensor needed to compute the emitted four-momentum is given by the sum of  the point-particle and tidal contributions, i.e.,
\begin{equation}
\label{Ttotal}
\tilde{T}^{\mu\nu} = \tilde{T}_{\rm pp}^{\mu\nu} + \tilde{T}_{\rm tid}^{\mu\nu} \, ,
\end{equation}
where the tilde denotes the  Fourier transform, $ \tilde T^{\mu \nu} (k) = \int d^4 x  \,  T^{\mu\nu}(x) e^{ i k \cdot x}$.
The stress-energy tensor in  the  point-particle case was computed in \cite{Mougiakakos:2021ckm,Riva:2021vnj} (see also \cite{Goldberger:2016iau}). Discarding the static contribution, which does not enter  the calculation, the leading-order diagrams are represented in Fig.~\ref{Figure}.
Using the notation
$\int_q \equiv \int \frac{d^4 q}{(2 \pi)^4}$ and  $\dd^{(n)} (x)  \equiv (2 \pi)^n \delta^{(n)} (x)$, it can be written as  
\be
\begin{split}
\label{Tmnpp}
& \tilde{T}^{\mu\nu}_{\rm pp}(k)   = \frac{m_1 m_2}{4 \mpl^2} \int_{q_1,q_2}  \!\!\!  \dd (q_1\cdot u_1) \dd (q_2\cdot u_2)\dd^{(4)}(k-q_1-q_2)   \\
& \  \times\frac{e^{i q_1 \cdot b_1+i q_2\cdot b_2}}{q_1^2q_2^2}   \left[ {t}^{\mu\nu}_{\firstc} (q_1 ,q_2) + {t}^{\mu\nu}_{\secondc}  (q_1 ,q_2)+ {t}^{\mu\nu}_{\vdash} (q_1 ,q_2) \right]\, ,
\end{split}
\ee
where ${t}^{\mu\nu}_{\firstc}$ is the contribution from diagram (a),  ${t}^{\mu\nu}_{\secondc}$ from the same diagram but with the two particles exchanged and ${t}^{\mu\nu}_{\vdash}$ from  (b).
We defer the reader to Ref.~\cite{Mougiakakos:2021ckm,Riva:2021vnj} for their explicit expressions.

The contribution of the tidal operators to the stress-energy tensor has no static piece. The leading PM term can be obtained from the diagram (c) in Fig.~\ref{Figure} and it is symmetric under exchange of the two particles. 
We obtain
\begin{align}
\tilde{T}_{\rm tid}^{\mu\nu}  = &\ \frac{m_1 m_2}{4 \mpl^2} \int_{q_1,q_2} \!\!\!  \dd (q_1\cdot u_1) \dd (q_2\cdot u_2)\dd^{(4)}(k-q_1-q_2) \nonumber \\
&   \times  \frac{e^{i q_1 \cdot b_1+i q_2\cdot b_2}}{q_1^2q_2^2} \sum_{ a =1,2} \sum_{ X = E,B}    t_{X_a^2}^{\mu\nu}(q_1, q_2)  \, ,
\label{Tmntidal}
\end{align}
with
$t_{X_1^2}^{\mu\nu}  \equiv - 2 \frac{c_{X_1^2}}{m_1}q_1^2 \eta^{\mu \alpha} \eta^{\nu \beta} \Pi^{{X}_1}_{ \alpha\beta,\kappa\lambda} \left( u_2^\kappa u_2^\lambda -  \eta^{\kappa \lambda}/2 \right)  $,
and an analogous formula for $(1 \leftrightarrow 2)$. 
The explicit   expressions  of $t_{X_1^2}^{\mu\nu}$ and $t_{X_2^2}^{\mu\nu}$ after use of eq.~\eqref{PiXa}, as well as the calculation of the waveform in direct space, are given in the Supplemental Material (SM).

Note that, because of scaling arguments, the tidal contribution vanishes in the soft limit $\omega \rightarrow 0$. As a consequence, at this order in $G$ tidal fields leave no trace on the gravitational wave memory.  (See the SM for details.) 
Since the emitted  angular momentum, $J^{\rm rad}$, at $\mathcal{O}(G^2)$ is proportional to the gravitational wave memory \cite{Damour:2020tta}, we  conclude that there are no  tidal effects on the emission of angular momentum at this order.  On the other hand, at leading-PM-order the radiation reaction on the scattering angle is related to the radiated angular momentum by $\chi^{\rm rad} = \frac12 \chi^{\rm cons}_{\rm LO}  J^{\rm rad}$ \cite{Bini:2012ji,Damour:2020tta,Bini:2021gat}, where the leading-order conservative contribution to the scattering angle, $\chi^{\rm cons}_{\rm LO}$, is of ${\cal O}(G)$. As a consequence, $\chi^{\rm rad} $ is unaffected by tidal effects  at ${\cal O}(G^3)$.


{\em Radiated four-momentum --}
The derivation of the emitted linear momentum  closely follows the procedure presented in Ref.~\cite{Riva:2021vnj}. 
 In particular, the emitted four-momentum is given as an integral over phase space of the outgoing graviton momentum $k^\mu$  weighted by  the probability of one graviton emission, which here is given  by the square of the  total stress-energy tensor from eq.~\eqref{Ttotal}. Although we use a quantum mechanical language, this quantity is well defined classically \cite{Goldberger:2007hy,Riva:2021vnj}.
Defining
$\dd_{\pm} (k^2) \equiv \theta(\pm k^0)\dd (k^2)$ we obtain,
for the {\em leading-order} contribution from the tidal effects to the radiated momentum, 
 \begin{equation}
P^\mu_{\rm tid}  = \frac{1}{2 \mpl^2} \int_k  \dd_+ (k^2) k^\mu  \text{Re}\left[\tilde{T}_{\rm pp}^{\alpha\beta}{\mathbb{P}_{\alpha\beta \rho\sigma} } \tilde{T}^*_{\rm tid}{}^{\rho\sigma}\right] \;.
\label{eq:Prad_Tidal}
\end{equation}  
 From the relation with tidal Love numbers below eq.~\eqref{linear}, for $R_a \sim G m_a$, the contribution quadratic in $\tilde{T}_{\rm tid} $ is further suppressed by $ {\cal O}(G^4)$ and is thus neglected.    

Following \cite{Riva:2021vnj}, we can interpreted the phase-space delta function as a cut propagator, so that the integrand reproduces a vacuum-to-vacuum diagram with a cut,
pictorially represented as
\be
\label{eq:2loop}
\begin{split}
&  \frac{\dd_+ (k^2)}{2 \mpl^2}   \text{Re}\left[\tilde{T}_{\rm pp}^{\alpha\beta}{\mathbb{P}_{\alpha\beta \rho\sigma} } \tilde{T}^*_{\rm tid}{}^{\rho\sigma}\right]  = \\
   {\raisebox{24pt}{\scalebox{0.7}{\MdiagT}}}  \!\!\! &+     {\raisebox{24pt}{\scalebox{0.7}{\NdiagT}} } \!\!\!   +    {\raisebox{24pt}{\scalebox{0.7}{\IYdiagT}}} \! + (1 \leftrightarrow 2) \;,
\end{split}
\ee 
where, using eqs.~\eqref{Tmnpp} and \eqref{Tmntidal} for the stress-energy tensor, the three topologies 
come  from considering the contributions from ${\rm Re} \big[{t}^{\alpha\beta}_{\firstc} {\mathbb{P}_{\alpha\beta}}^{\rho\sigma} t^{{X}_1 \,*}_{\rho\sigma} \big]$, ${\rm Re} \big[{t}^{\alpha\beta}_{\secondc} {\mathbb{P}_{\alpha\beta}}^{\rho\sigma} t^{{X}_1 \,*}_{\rho\sigma} \big]$ and ${\rm Re} \big[{t}^{\alpha\beta}_{\vdash} {\mathbb{P}_{\alpha\beta}}^{\rho\sigma} t^{{X}_1 \,*}_{\rho\sigma} \big]$,
respectively. Notice that the $H$ diagram is absent, because there are no tidal interactions linear in $h_{\mu \nu}$.

We can now recast the problem of computing the emitted momentum 
as evaluating
a cut 2-loop integral followed by a 2d Fourier transform. In particular,  the emitted four-momentum 
can be 
decomposed  without loss of generality along $\check{u}_1^\mu \equiv (u_1^\mu - \gamma u_2^\mu)/(1-\gamma^2)$, $\check{u}_2^\mu \equiv (u_2^\mu - \gamma u_1^\mu)/(1-\gamma^2)$ (satisfying $u_a \cdot \check{u}_b = \delta_{ab}$), with $\gamma \equiv u_1 \cdot u_2$, and $b^\mu$. By the symmetries of the integrand, one can show  that the component along $b^\mu$ vanishes, so that the momentum can be written as
\be 
\label{eq:PTrad_E}
P^\mu_{\text{tid}}  =  \frac{15  \pi G^3 m_1^2 m^2_2}{64 |\vb|^7} \sum_{X} \bigg[  \frac{c_{X_1^2}}{m_1} \Big( {\cal E}^X  \check{u}_1^\mu  + {\cal F}^X  \check{u}_2^\mu  \Big)  + (1 \leftrightarrow 2) \bigg]\, .
\ee
The functions ${\cal E}^X$ and ${\cal F}^X$ inside the brackets depend only on $\gamma$
and can be expressed as
2d Fourier transforms of cut 2-loop integrals,  ${\cal I}_{{\cal E}}^{{X}}(\gamma)$ and ${\cal I}_{{\cal F}}^{{X}}(\gamma)$. For instance,  
\begin{equation}
\label{eq:calEX}
\begin{split}
 {\cal E}^{X} (\gamma) \!=\! |\vb|^7   \!\!\!\int_q & \dd (q\cdot u_1) \dd (q\cdot u_2) e^{i q\cdot b}   (-q^2)^{5/2} {\cal I}_{{\cal E}}^{{X}}(\gamma) \, ,
\end{split}
\end{equation}
and 
analogously
for ${\cal F}^{X} (\gamma)$.
The explicit expressions of the 2-loop integrals are given in the SM.

Making use of reverse unitarity \cite{Anastasiou:2002yz,Anastasiou:2002qz,Anastasiou:2003yy,Anastasiou:2015yha}, we can  
use IBP identities to express the 2-loop integrals $ {\cal I}_{{\cal E}, {\cal F}}^{{X}}$  as linear combinations of   simpler master integrals.
We perform this reduction using the \texttt{Mathematica} package \texttt{LiteRed} \cite{Lee:2012cn, Lee:2013mka}, finding that 
the  three integrals defined in eqs.~(4.13)--(4.15) of Ref.~\cite{Riva:2021vnj}
form a complete base. (In the minimally coupled case we need a fourth integral, defined in eq.~(4.16) of this reference. This comes from the $H$ diagram, which here is absent.)
These integrals can be solve using differential equation methods \cite{Kotikov:1990kg,Bern:1992em,Gehrmann:1999as,Henn:2013pwa, Caron-Huot:2014lda,Henn:2014qga,Parra-Martinez:2020dzs}. 
Eventually, we find that
\be 
 {\cal E}^{X}  =  f_1^{{X}}   + f_2^{{X}}   \log\left(\frac{\gamma + 1}{2}\right) + f_3^{{X}}   \frac{  \text{arccosh}  (\gamma) }{\sqrt{\gamma^2-1}} \;,
\ee
with $f_1^{{X}}$, $f_2^{{X}}$, $f_3^{{X}}$ and  ${\cal F}^{X}$ given in Table~\ref{table}.
\begin{table}[tb]
	\setlength{\tabcolsep}{1pt} 
	\renewcommand{\arraystretch}{3}
	\begin{tabular}{|c|}
		\hline
		\scalebox{0.84}{\tabeq{10cm}{
f^{E}_{1} & = \frac{1}{2 (\gamma +1)^3 \sqrt{\gamma ^2-1}}  \Big[937\gamma ^9 + 1551 \gamma ^8 -2463 \gamma ^7 - 5645 \gamma ^6 \\ 
& + 20415 \gamma^5 + 65965 \gamma^4 - 349541 \gamma^3 + 535057 \gamma^2 - 360356 \gamma +92160 \Big]\,  \\
f^{B}_{1} & =   \frac{\gamma-1}{4 (\gamma +1)^3 \sqrt{\gamma ^2-1}}  \Big[  1559 \gamma ^8 +3716  \gamma ^7 -1630 \gamma ^6  - 11660 \gamma^5  \\ 
& - 28288 \gamma^4 + 155292 \gamma^3 - 543442 \gamma^2 + 535212 \gamma - 180775 \Big]\,  \\
f^{E}_{2}  & =  {30 \sqrt{\gamma ^2-1} \left(21 \gamma ^4-14 \gamma ^2+9 \right)} \, , \\ 
f^{B}_{2}   & =   210 (\gamma^2 -1)^{3/2} \left( 1+3 \gamma^2 \right)  \; , \qquad
f^{X}_{3}  = - {f^{X}_{2}} \frac{\gamma(2 \gamma^2 -3)}{4 (\gamma^2-1)}   \,   \\
{\cal F}^E  & =   \frac{ 3 (\gamma - 1)^2  }{ (\gamma +1)^3 \sqrt{\gamma ^2-1} }  \Big[ 42 \gamma ^8 +210 \gamma ^7 + 315 \gamma ^6  -105 \gamma^5 - 944 \gamma^4 \\
& - 1528 \gamma^3 +22011 \gamma^2 -33201 \gamma +16272\Big] \,  \\
{\cal F}^B &  = -  \frac{3 (\gamma-1)^3  (105 \gamma^5  + 630 \gamma^4 \!+\!1840 \gamma^3 \!+ 3690 \gamma^2 \!- \! 17769 \gamma + 15984) }{ (\gamma +1)^3 \sqrt{\gamma ^2-1}}  \, 		}}
\\
		\hline
	\end{tabular}
    \caption{Functions specifying the radiated four-momentum in eq.~\eqref{eq:PTrad_E}.}
	\label{table}
\end{table}

From eq.~\eqref{eq:PTrad_E}, one can compute the   radiated energy in the center-of-mass  frame 
from
tidal effects, $\Delta  E_{\rm tid}  \equiv P_{\rm tid} \cdot u_{\rm c.m.}$. Defining  the total mass $m \equiv m_1+m_2$, the symmetric mass ratio $\nu \equiv m_1 m_2/m^2$ and $h (\nu,\gamma) \equiv E  /m = \sqrt{1+2 \nu (\gamma-1)}$, where $ E  $ is the incoming energy of the two-body system, this reads
\be 
\label{ECoM}
 \Delta E_{\rm tid}  =  \frac{15  \pi G^7 m^8 \nu^2}{64 |\vb|^7 h }  {\cal G} ({\cal E}^X, {\cal F}^X )   \, ,
\ee
 where 
\be
{\cal G} ({\cal E}^X, {\cal F}^X ) \equiv  \sum_{X}   \left[ \kappa_{X^2} {\cal E}^X \!+  \lambda_{X^2}  ({\cal F}^X  - {\cal E}^X)  \right] \, ,
\ee
and we have introduced the dimensionless parameters \cite{Kalin:2020lmz}
\begin{align}
\lambda_{X^2} &\equiv \frac{1}{G^4 m^5 } \left(  \frac{c_{X_1^2} m_2}{m_1}   +  \frac{c_{X_2^2} m_1}{m_2} \right)\; , \\
\kappa_{X^2} &\equiv  
\frac{1}{G^4 m^4 } \left(  \frac{c_{X_1^2}}{m_1}     + \frac{c_{X_2^2}}{m_2} \right) \;.
\end{align}
Expanding for small velocities $v \equiv \sqrt{\gamma^2 -1 }/\gamma$, we find 
\be
\begin{split}
{\cal E}^E & = 288 v^3 + \frac{2143}{7} v^5 + \frac{14542}{21} v^7 + {\cal O}(v^9) \;, \label{EEsmallv}\\
{\cal E}^B & = -98 v^5 + \frac{585}{4} v^7 + {\cal O}(v^9) \;, \\
{\cal F}^E & = 288 v^3 + 336 v^5 + \frac{3027}{4} v^7 + {\cal O}(v^9) \;, \\
{\cal F}^B & = -210 v^5 - \frac{669}{4} v^7 + {\cal O}(v^9) \;,
\end{split}
\ee
which shows that the current (magnetic)  quadrupole is 1PN order higher than the mass (electric) one, as expected.

Finally, the emitted energy from a two-body encounter can be used to    derive the energy loss for  
closed
orbits by the use of the B2B relation \cite{Kalin:2019rwq,Kalin:2019inp,Bini:2020hmy,Cho:2021arx},
$\Delta E ^{\rm (closed)} (\gamma, J) = \Delta  E ^{\rm (open)} (\gamma, J)  - \Delta E ^{\rm (open)} (\gamma, - J)$, 
where the emitted energy on the right-hand side must be expressed in terms of the angular momentum $J=|\vb| m \nu \sqrt{\gamma^2- 1}/h (\nu,\gamma)$ (with $h= E/m$) and analytically continued to bound orbits 
with $h<1$,  corresponding to $\gamma = 1 +  \frac{h^2 - 1 }{2 \nu } < 1$.
This yields
\be
\label{EAC}
\Delta  E^{\rm (closed)}_{\rm tid}  = \frac{15  \pi G^7 m^{15} \nu^9 (1- \gamma^2)^{7/2}}{64 J^7 h^8 }  {\cal G} ( \tilde {\cal E}^X, \tilde {\cal F}^X ) \\
\ee
where 
\begin{align}
	\tilde {\cal E}^X & = \tilde{f}_1^{{X}}   + \tilde{f}_2^{{X}}   \log\left(\frac{\gamma + 1}{2}\right) + \tilde{f}_3^{{X}}   \frac{  \text{arccos}  (\gamma) }{\sqrt{1-\gamma^2}} \, , 
	\end{align}
with $\tilde{f}_{1}^X = -2  f_{1}^X$, $\tilde{f}_{2}^E = -2  f_{2}^E$, $\tilde{f}_{2}^B = 2  f_{2}^B$, $\tilde{f}_{3}^E = 2  f_{3}^E$, $\tilde{f}_{3}^B = -2  f_{3}^B$ and $\tilde {\cal F}^X = -2  {\cal F}^X$, all subject to the replacement $(\gamma^2 - 1)^{n/2} \to (1 - \gamma^2 )^{n/2}$. In the following we show that this expression is consistent with known results in the PN approximation.

 
{\em Radiated Flux --} 
The instantaneous flux is defined as $F \equiv dE/dt$. Focussing on the tidal correction, $F_{\rm tid}$, and integrating this relation for half of the scattering trajectory, we obtain
\begin{equation}
	 \Delta E_{\rm tid}  (\gamma) = 2\int_{|\vb|}^{\infty}\! \frac{dr}{\dot r}  F_{\rm tid}(r, \gamma)  \, .
	 \label{eq:EnFlux}
\end{equation}
We have assumed that the expression of the flux is in isotropic gauge; thus, we have dropped the dependence on $J$ in $F_{\rm tid}$.
From eq.~\eqref{ECoM}, the leading-order tidal contribution to the flux scales as $G^7$ so that its dependence on $r$ is fully determined:
$	F_{ {\rm tid}}(r, \gamma) \propto r^{-8} $.
By integrating the right-hand side of eq.~\eqref{eq:EnFlux} with this ansatz, and using $\dot r$ for straight orbits at this PM order, we find  
\begin{equation}
	F_{{\rm tid}} (r, \gamma) =  \frac{G^7 m^8}{r^8}  \frac{3\nu^3\sqrt{\gamma^2 - 1}}{4 h^3 \xi} {\cal G} ({\cal E}^X, {\cal F}^X )   \, ,
	\label{eq:PMflux}
\end{equation}
where $\xi \equiv E_1 E_2/E^2$, and $E_a$ is the initial asymptotic energies of body $a=1,2$.
This result extends the one for point-particles  computed in \cite{Cho:2021arx}. As discussed there, due to the absence of a  term higher in $G$, the leading PM computation is insufficient to reconstruct the  leading PN flux but it provides the full velocity--or reduced-energy--series to order $G^3$.


{\em Consistency check --} 
 We can compare our result for small velocities to the emitted flux and energy  in one period derived in the PN expansion in the large eccentricity limit, i.e.~to leading order in large $J$. 

The tidal effects on the gravitational wave energy flux   for spinless bodies has been computed up to the next-to-next-to-leading PN order in \cite{Henry:2020ski} (see  \cite{Henry:2019xhg,Henry:2020pzq} for a derivation of the equations of motion and Hamiltonian in this case, respectively; see also \cite{Huber:2020xny} for a calculation of the PM Hamiltonian and the emitted energy for quasi-circular orbits at leading PN order, with interactions cubic in the curvature and tidal effects). Although in that reference the results were given only for quasi-circular orbits, their authors have kindly provided us with an expression  of the flux  $F_{\rm tid}^{\rm (PN)}$ and the conserved energy $ E $ and angular momentum $J$
for generic orbits, written in terms  of $r$, $\dot r$ and $\dot \phi$, respectively the two-body distance, the radial velocity and the angular velocity in the center-of-mass frame.
We have used the expressions for $ E $ and $J$ to  replace $\dot r = \dot r(r,  E ,J)$ and $\dot \phi=\dot \phi (r,  E ,J)$ in the flux and we have computed the emitted energy {\em for generic 
closed 
orbits} by integrating it in the variable $r$ over one period. 

The resulting energy reduces to that given in \cite{Henry:2020ski} for circular orbits. Moreover,  it is consistent with the expansion eq.~\eqref{EEsmallv}  (taking
into account the factor of -2 according to eq.~\eqref{EAC}). Since all the powers of $\gamma$ in Table~\ref{table} intervene in this expansion, this is a rather nontrivial check of our calculation.
 Moreover, the PN flux  $F_{\rm tid}^{\rm (PN)}$ coincides with the low-velocity expansion of eq.~\eqref{eq:PMflux}, up to total derivatives in the balance equations--the so-called Schott terms. Although the two fluxes are written in different gauges (in harmonic and isotropic gauge, respectively in Ref.~\cite{Henry:2020ski}   and in eq.~\eqref{eq:PMflux}) the gauge difference is  2PM orders higher and can be neglected. 
For the reader's convenience, we  report the   explicit expression of the PM flux  in the ancillary file submitted with the arXiv version of this article.
 


{\em High-energy limit --} Going back to the energy loss for hyperbolic-like orbits, eq.~\eqref{ECoM}, for large $\gamma $ we find   
${\cal E}_{\rm HE}^X= (a_X + b_X \log \gamma) \gamma^5 + {\cal O}(\gamma^{3})$ and ${\cal F}_{\rm HE}^X= c_X \gamma^6 + d_X \gamma^4  + {\cal O}(\gamma^{2})$, with $a_E=937/2-945 \log 2$, $a_B=1559/4-945 \log 2$, $b_E=b_B=315$, $c_E=126$, $c_B=0$, $d_E=-504$ and $d_B=-315$. 
While ${\cal E}_{\rm HE}^E$ and ${\cal E}_{\rm HE}^B$ scale in the same way with $\gamma$,  ${\cal F}_{\rm HE}^E$ and ${\cal F}_{\rm HE}^B$  behave differently. Our perturbative expansion is valid for $\gamma  (G m /|\vb|) \ll 1$ \cite{DEath:1976bbo,Damour:2019lcq,DiVecchia:2022nna} (see also \cite{Kovacs:1978eu}). In this regime $\Delta E_{\rm tid} \ll \Delta E \sim (Gm/|\vb|)^3 (m/h) \gamma^3 \ll E$. 
%


{\em Conclusion --} We have computed the four-momentum  and the flux emitted in gravitational waves by the scattering of  tidally interacting bodies at leading order in the  post-Minkowskian approximation. Our computation uses the worldline effective field theory  approach and the results obtained are, up to our knowledge, new. We focused on electric and magnetic-type quadrupolar effects but our computations can be straightforwardly extended to higher multipoles or to higher-orders in the curvature fields. 

We 
have derived the
emitted energy for bound orbits using the B2B dictionary and 
verified that it is consistent 
with PN results for eccentric orbits. 
Considering the ultra-relativistic limit of the energy loss, we observe that the contributions of the electric and magnetic component scale differently unlike the case of the conservative scattering angle. 
It would be interesting to use the  derived PM flux to study the corresponding modifications of the waveform.

\vspace{0.4cm}


{\em Acknowledgements --} We  thank Luc Blanchet, Guillaume Faye and Quentin Henry  for helpful discussions and for kindly providing us with expressions of the  flux, conserved energy and angular momentum for generic orbits. We also thank  Thibault Damour, Carlo Heissenberg, Enrico Herrmann, Gregor K\"alin, Alessandro Nagar, Rafael Porto, Rodolfo Russo and Leong Khim Wong for interesting discussions and comments on the draft.  
This work was partially supported by the CNES. 

\onecolumngrid

\vspace{0.3cm}

\begin{center}
\Large{\bf Supplemental Materials}
\end{center}
\twocolumngrid


\section*{ Tidal stress-energy tensor} 
Here we give the explicit expression of the tidal stress-energy tensor defined in eq.~\eqref{Tmntidal}. Introducing $\beta \equiv 2 \gamma^2 -1$ and $\omega_a=k\cdot u_a$, we have
\begin{align}
&\frac{m_1}{q_1^2} \frac{t_{E_1^2}^{\mu\nu}}{c_{E_1^2} }=\omega_1^4\eta^{\mu\nu}-\beta \omega_1^2  q_{2}^{\mu}q_{2}^{\nu} -2\omega_1^3q_{1}^{(\mu}u_{1}^{\nu)} \nonumber\\
&-\omega_1\left( \beta q_1^2+4\gamma\omega_1\omega_2+2 \omega_1^2\right)q_{2}^{(\mu}u_{1}^{\nu)}+4\gamma\omega_1^3q_{2}^{(\mu}u_{2}^{\nu)}  \nonumber\\
&-\left( \beta {q_1^4}/{4}+2 \gamma\omega_1\omega_2q_1^2+2 \omega_1^2\omega_2^2\right)u_{1}^{\mu}u_{1}^{\nu}\nonumber\\
& -2\omega_1^4u_{2}^{\mu}u_{2}^{\nu}+2\omega_1^2\left(\gamma {q_1^2}+2\omega_1\omega_2\right)u_{1}^{(\mu}u_{2}^{\nu)} \label{tid1} \\
&\frac{m_1}{q_1^2} \frac{t_{B_1^2}^{\mu\nu}}{c_{B_1^2} } =\omega_1^2( {q_1^2}/{2}+\omega_1^2)\eta^{\mu\nu}-\omega_1\left( \beta \omega_1- 2\gamma\omega_2\right)q_{2}^{\mu}q_{2}^{\nu}\nonumber\\
&+\omega^2_1q_{1}^{(\mu}q_{2}^{\nu)}-\frac{1}{2}(\gamma  {q_1^2} + 2\omega_1\omega_2)^2u_{1}^{\mu}u_{1}^{\nu}-\omega_1^2( {q_1^2}+2 \omega_1^2)u_{2}^{\mu}u_{2}^{\nu}\nonumber\\
&+ ( {q_1^2}/{2}+2\omega_1^2)(\gamma {q_1^2}+2\omega_1\omega_2) u_{1}^{(\mu}u_{2}^{\nu)}
-\left(2 \omega_1(\omega_1^2-\omega_2^2 + \right. \nonumber\\
&\left.+2\gamma\omega_1\omega_2)+{q_1^2}((1+4\gamma^2)\omega_1-2\gamma\omega_2)/2\right)q_{2}^{(\mu}u_{1}^{\nu)} +\omega_1(\gamma {q_1^2} \nonumber\\
&+4\gamma\omega_1^2- 2 \omega_1\omega_2)q_{2}^{(\mu}u_{2}^{\nu)} -\omega_1( {q_1^2}/{2}+2\omega_1^2)q_{1}^{(\mu}u_{1}^{\nu)} \, , \label{tid2}
\end{align}
and analogous expressions for $t_{X_2^2}^{\mu\nu}$ with $(1\leftrightarrow 2)$.


\section*{Waveform}
The  
asymptotic waveform in direct space, i.e.~$h_\lambda \equiv \epsilon^{\mu \nu}_\lambda h_{\mu \nu}/\mpl$, with $\lambda=\pm$, evaluated at distances $r$ much larger than the interaction region, reads 
\begin{align}
\label{waveform}
h_\lambda (x) &= \frac{4G}{  r}   \int \frac{d k^0}{2 \pi} e^{-i k^0 u}   \epsilon^{\lambda *}_{\alpha \beta} \tilde{T}^{\alpha\beta} |_{k^\mu = k^0 n^\mu} \;,
\end{align}
where $u \equiv t -r$ is the retarded time. The stress-energy tensor on the right-hand side is evaluated on-shell, i.e.~$k^\mu = k^0 n^\mu$, with $n^\mu \equiv (1, \vn)$ and 
$\vn^2=1$. 

In the point-particle case, the waveform   was computed using our approach in \cite{Mougiakakos:2021ckm} (see also \cite{Jakobsen:2021smu}), in agreement with earlier calculations \cite{Thorne:1975aa}.
Here we focus on the tidal contribution, obtained by replacing \eqref{Tmntidal} in the above expression.
Integrating first in $k^0$,  we obtain
\begin{align}
h_{\lambda}^{\rm tid}=& \!\!\! \sum_{X={E,B}} \!\!\! c_{X^2_ 1}\frac{Gm_2}{r\, \mpl^2}\frac{ \epsilon^{ * \lambda}_{\mu \nu} (\vn)}{n\cdot u_1}\int_q\frac{\dd(q\cdot u_2) e^{i q\cdot \tilde{b}} }{q^2 (k-q)^2}  \nonumber \\
& \times  {t_{X_1^2}^{\mu\nu}} (q-k,q) \big|_{k^0=\frac{q\cdot u_1}{n\cdot u_1}} +(1 \leftrightarrow 2),
\end{align}
where $\tilde{b} \equiv b+\frac{u_1}{n\cdot u_1}(u+\vn\cdot \vb_1)$. We can then choose a frame and remove the remaining delta function by integrating in $q^0$. Regardless of the chosen frame, the remaining integral can be 
put in the form of the master integral
\begin{equation}
I=\int_{\bf{q}}\frac{e^{i\bf{q}\cdot\bf{b}}}{{\bf q} \cdot {\bf M} \cdot {\bf q}}= \frac{1}{4 \pi}   \frac{\left[ {\bf b} \cdot {\bf M}^{-1} \cdot { \bf b} \right]^{-1/2}}{ \left[\text{det}(\bf{M})\right]^{1/2}} \; ,
\end{equation}
where ${\bf M}$ is a $3\times 3$ matrix, or of its tensorial generalization, $I^{i_1,...,i_n}=\int_{\bf{q}}\frac{{\bf q}^{i_1} \cdots {\bf q}^{i_n} e^{i\bf{q}\cdot\bf{b}}}{{\bf q} \cdot {\bf M} \cdot {\bf q}} $, which can be 
solved
by taking derivatives of the master integral,  $I^{i_1,...,i_n}=\frac{\partial}{i\partial b_{i_1}}...\frac{\partial}{i\partial b_{i_n}}I$.

Performing the calculation in the rest frame of particle 2, i.e.~for $u_1^{\mu}=\gamma(1, v\hat{\vv})$ and $u_2^{\mu}=\delta^{\mu}_0$, and choosing $b_1^{\mu}=(0,\vb )$ and $b_2^{\mu}=0$ for simplicity,
we obtain
\begin{equation}
h_{\lambda}^{\rm tid}=120\frac{G^2m_1m_{2}}{r}\frac{\gamma^2v^2}{|\vb|^5}\sum_{a=1,2}\sum_{X={E,B}}\frac{c_{X_a^2}}{m_a} \frac{\epsilon^{ * \lambda}_{ij}  \ve_I^i\ve_J^jA^{IJ}_{X_{a}^2}}{(n\cdot u_{a})^3 {c}_{a}^{9/2}}  \, ,
\label{wavefinal}
\end{equation}
where we have defined $\ve_I \equiv (\hat \vv,  \hat \vb)$ (with $I=v,b$), and the functions ${c_1}=1+ \frac{\gamma^2 v^2}{ (n\cdot u_1)^2}\big(\frac{u}{|\vb|}+\vn \cdot \hat{\vb} \big)^2$ and ${ c_2}=1+\gamma^2 v^2 \frac{u^2}{|\vb|^2}$. Defining $f_a=6 c_a - 7$, $g_a=\sqrt{c_a - 1} (4 c_a - 7)$, the explicit expressions for $A^{IJ}_{X_{a}^2}$  are
\begin{align}
&A^{bb}_{ {E}_{a}^2}={(n\cdot u_a)^2} \beta f_a \;, \nonumber\\
&A^{vb}_{ {E}_{a}^2}=\gamma(n\cdot u_a)\big[ f_a \beta  v \delta_{a1} \hat{\vb} \cdot \vn +g_a\big] \;, \nonumber\\
&A^{vv}_{ {E}_{a}^2}=\frac{ (2f_a^2-17f_a-7)}{30}+f_a(\gamma^2-1)  \nonumber\\
& +   v \gamma^2 \delta_{a1} \hat{\vb} \cdot \vn \big[f_a \beta  v \delta_{a1} \hat{\vb} \cdot\vn+2g_a\big] \;, \nonumber\\
&A^{bb}_{ {B}_{a}^2}=2\gamma{(n\cdot u_a)}\big[\gamma\ n\cdot u_a-\frac{(n\cdot u_1)(n\cdot u_2)}{n\cdot u_a}\big]f_a \;, \nonumber\\
&A^{vb}_{ {B}_{a}^2}=f_a\gamma^2(2\gamma\delta_{a1} n\cdot u_a - 1)v \hat{\vb} \cdot \vn \nonumber\\
&+g_a\big[\gamma\ n\cdot u_a-\frac{(n\cdot u_1)(n\cdot u_2)}{n\cdot u_a}\big]\;, \nonumber\\
&A^{vv}_{ {B}_{a}^2}=2\gamma^2v \hat{\vb} \cdot \vn\big[ f_a\gamma^2v \delta_{a1}\hat{\vb} \cdot \vn +\frac{\gamma\delta_{a1}-\delta_{a2}}{\gamma}g_a\big] \;.
\end{align}
One can verify that, upon PN expanding, the contribution of the current (magnetic) quadrupole enters at 1PN higher than the mass (electric) one, as expected.

Note, also, that the gravitational wave memory, i.e., the difference in  the waveform  between asymptotic past and future, defined as 
\begin{equation}\label{memory}
\Delta h_{\lambda} ({\bf x}) \equiv\int_{-\infty}^{+\infty}du\ \dot{h}_{\lambda} (u , {\bf x})\;, 
\end{equation}
is not affected by tidal deformations at this order in $G$. Indeed, from eq.~\eqref{waveform} the contribution of tidal effects to the memory reads
\begin{equation}
\Delta h^{\rm tid}_{\lambda}= -i\frac{4G}{  r}   \int  \frac{d k^0}{2 \pi}   \dd(k^0) k^0 \, \epsilon^{\lambda *}_{\alpha \beta} \tilde{T}^{\alpha\beta}_{\rm tid} |_{k^\mu = k^0 n^\mu} \;.
\end{equation}
From eqs.~\eqref{tid1} and \eqref{tid2} above,  $\tilde{T}_{\rm tid}^{\mu\nu}\sim \mathcal{O}((k^0)^4)$ and thus the right-hand side of this equation vanishes. This conclusion can be extended to higher-order tidal fields, which contribute to $\tilde{T}_{\rm tid}$ higher in $k^0$.

\section*{2-loop integrals}  

Introducing the following basis of propagators
\begin{align}
\rho_1 & = 2 \ell_1 \cdot u_1\, ,  &  \rho_2 & = - 2 \ell_1 \cdot u_2\, , &  \rho_3 & = - 2 \ell_2 \cdot u_1\, \notag\\
\rho_4 &= 2 \ell_2 \cdot u_2 \, ,& \rho_5  &= \ell_1^2 \, ,  & \rho_6 &= \ell_2^2 \, ,  \\
\rho_7 &= (\ell_1 + \ell_2 - q)^2 \, , &  \rho_8 &= (\ell_1 - q)^2 \, ,  & \rho_9 &= (\ell_2 - q)^2 \, ,  \notag
\end{align}
we can explicitly write the 2-loop integrals in eq.~\eqref{eq:calEX}~as
\begin{align}
{\cal I}^{X}_{\cal E} & = \frac{2^{17} \pi^2}{15 (-q^2)}\frac{m_1}{c_{X_1^2}} \int_{\ell_1, \ell_2} \!\!\!\!\!\dd_{-}(\rho_7)\dd(\rho_1)\dd(\rho_4) \frac{\rho_3 \, {\cal N}_{X_1}}{\rho_5\rho_6\rho_8\rho_9} \,, \\
{\cal I}^{X}_{\cal F} & = \frac{2^{17} \pi^2}{15 (-q^2)}\frac{m_2}{c_{X_2^2}} \int_{\ell_1, \ell_2} \!\!\!\!\!\dd_{-}(\rho_7)\dd(\rho_1)\dd(\rho_4) \frac{\rho_3 \, {\cal N}_{X_2}}{\rho_5\rho_6\rho_8\rho_9} \,,
\end{align}
where 
${\cal N}_{X_a}  \equiv {\rm Re} \left[ \left( {t}^{\alpha\beta}_{\firstc} + {t}^{\alpha\beta}_{\secondc} + {t}^{\alpha\beta}_{\vdash} \right){\mathbb{P}_{\alpha\beta}}^{\rho\sigma} t^{{X}_a \,*}_{\rho\sigma} \right] $.

\vfill


\bibliography{ref}
\bibliographystyle{utphys}

\end{document}